\documentclass[preprintnumbers,superscriptaddress,pra,showkeys]{revtex4}
\usepackage{amsfonts}
\usepackage{amssymb}
\usepackage{amsmath}
\usepackage{epsfig}
\usepackage{graphicx}
\usepackage{color}

\input{tcilatex}
\begin{document}

\title{Asynchronous finite differences in\ most probable distribution\emph{\ 
}with finite numbers of particles}
\author{Q. H. Liu}
\email{quanhuiliu@gmail.com}
\affiliation{School for Theoretical Physics, School of Physics and Electronics, Hunan
University, Changsha 410082, China}
\date{\today }

\begin{abstract}
For a discrete function $f\left( x\right) $ on a discrete set, the finite
difference can be either forward and backward. If $f\left( x\right) $ is a
sum of two such functions $f\left( x\right) =f_{1}\left( x\right)
+f_{2}\left( x\right) $, the first order difference of $\Delta f\left(
x\right) $ can be grouped into four possible combinations, in which two are
the usual synchronous ones $\Delta ^{f}f_{1}\left( x\right) +\Delta
^{f}f_{2}\left( x\right) $ and $\Delta ^{b}f_{1}\left( x\right) +\Delta
^{b}f_{2}\left( x\right) $, and other two are asynchronous ones $\Delta
^{f}f_{1}\left( x\right) +\Delta ^{b}f_{2}\left( x\right) $ and $\Delta
^{b}f_{1}\left( x\right) +\Delta ^{f}f_{2}\left( x\right) $, where $\Delta
^{f}$ and $\Delta ^{b}$ denotes the forward and backward difference
respectively. Thus, the first order variation equation $\Delta f\left(
x\right) =0$ for this function $f\left( x\right) $ gives at most four
different solutions which contain both true and false one. \emph{A formalism
of the discrete calculus of variations is developed to single out the true
one by means of comparison of the second order variations, in which the
largest value in magnitude indicates the true solution, yielding the exact
form of the distributions for Boltzmann, Bose and Fermi system without
requiring the numbers of particle to be infinitely large}. When there is
only one particle in the system, all distributions reduce to be the
Boltzmann one.
\end{abstract}

\keywords{statistical distribution, discrete calculus of variations, most
probable distribution}
\maketitle

\section{Introduction}

This paper aims at solving a long time and also fundamental problem in
statistical physics. For the historical side which was the initial
motivation of the present study, we are familiar with a difficulty of
possible existence of a derivation of Bose-Einstein (or simply Bose),
Fermi-Dirac (or simply Fermi) and Boltzmann-Maxwell (or simply Boltzmann)
distribution without use of the Stirling approximation of the factorials,
within the method of\emph{\ most probable distribution }(MPD) in statistical
physics. Some feels lukewarm with the use of the Stirling approximation, and
some thinks it a serious defect, but no one considers it comfortable. The
comments of Tolman in 1938 on the problem are worthy of mention. \cite%
{Tolman} The first part of the comments is that, with the help of the
approximation, we can use the calculus of continuous functions instead of
that of the discrete ones. The original sentences include: "To carry out the
analysis, let us take the numbers of molecules $n_{i}$ ... as being large
enough, not only to permit the foregoing use of Stirling's approximation for
their factorials, but also to justify us in treating the numbers themselves
as continuous variables in applying the calculus of variations." (p.79), and
"This was done in order to employ the usual methods of the calculus of
variations as applied to continuous variables." (p.80). The second part of
the comments is that, the approximations are evidently worse sometimes. The
original sentences include: "And, from a computational point of view, it
will be remembered that our previous calculations of the molecular
distributions to be expected at equilibrium were actually carried out for
simplicity, with a somewhat unsatisfactory introduction of the Stirling
approximation for factorial numbers, which, as emphasized by Fowler, might
even involve the use of that approximation for integers as small as zero or
one." (p. 481), and "The derivations which we have given for these relations
have been obtained with the help of the Stirling approximation for
factorials, and this may be regarded to some extent as a defect, ... ."
(p.373). The similar comments can also be found in vast literature, among
them I like to mention two \cite{wang,mpd}. For the present side which urges
the present study, we are impressed by the recent experimental realization
of the single-atom heat engine \cite{single}. The single particle exists in
a thermal distribution whose width is proportional to its temperature, which
clearly is the Boltzmann distribution \cite{single}. We know that in the
classical limit, all Bose, Fermi and Boltzmann distributions are all reduced
to the Boltzmann one, but whether it is so in the limit of single atoms
poses a contemporary problem.

The method of MPD\emph{\ }is the most common way used to derive various
statistical distributions in mathematics, physics, chemistry, materials
science and computational science, etc. The functions in statistics are
usually are defined on a discrete lattice rather than a continuous interval,
so we must be able to deal with the differences, difference quotients and
sums of discrete functions, \cite{arnold,Goldstein} instead of the
differentials, derivatives and integrations of the continuous functions.
However, once manipulating the discrete calculus, we immediately run into a
problem: Given a discrete function or functional, the difference can be
forward, backward, central, and even more complicated combinations of these
differences, so that the first order derivatives or variationals lead to
many possibilities which can not be all true. In statistical physics, we are
familiar with Boltzmann, Bose and Fermi system of many particles, in which
there are a huge amount of distributions compatible with the constraint
conditions. To obtain the MPD, the routine manner is to render the discrete
problem into a continuous one with use of the Stirling approximation of the
factorials, which is more accurate as the variables of factorials are
larger. \cite{Tolman,wang,mpd} In the thermodynamic limit of an infinitely
large number of particles, the obtained\ MPD is the true one. However, the
fetal shortcoming of the routine manner lies in that the variables of
factorials are frequently small, and they can even be one or two. To note
that in mathematics, the Stirling approximation can be accurate when the
variables can be very small, but in statistical physics, we always mean that 
$\ln n!\approx n\left( \ln n-1\right) $. \cite{Tolman,wang,mpd} It is
understandable for the exact form of distributions for Boltzmann, Bose and
Fermi system of finite numbers of particles has posed a formidable problem
for long time. \cite{Tolman,wang,mpd,chem,mermin,india,PLA,amj} Nowadays,
the statistical mechanics for finite number of particles attracts much
attention. \cite{single,finite1,finite2,finite3,finite4} In present work, I\
report an exact discrete calculus of variations to give the exact
distributions for the Boltzmann, Bose and Fermi system of finite numbers of
particles.

In section II, we present a new procedure of discrete calculus of
variations. In section III and IV, by use of this procedure, we show how to
derive the Bose, Fermi and Boltzmann distribution without invoking the
Stirling approximation, where the section IV focuses on the distributions
with a few number of particles. In final section V, a brief conclusion is
given.

\section{A new discrete calculus of variations}

At first, given a\emph{\ function} $f\left( x\right) $ of variable $x$ on a
discrete lattice. The first order finite \emph{differences }of the function%
\emph{\ }$f\left( x\right) $ can be grouped into two groups. \cite%
{arnold,Goldstein} The first group is the \emph{forward} difference: $\left(
\Delta ^{f}\right) f\left( x\right) =f\left( x+h\right) -f\left( x\right) $ (%
$h>0$), and\emph{\ the second group is the backward} difference $\left(
\Delta ^{b}\right) f\left( x\right) =f\left( x\right) -f\left( x-h\right)
\neq \left( \Delta ^{f}\right) f\left( x\right) $. This positive and finite
constant $h$ has different meanings in different situations. In numerical
calculations of continuous functions, $h$ may be determined by the
computational accuracy. For the discrete function, $h$ can not be
arbitrarily chosen, and especially in discrete calculus of variationals, $h$
can only take the smallest distance between two nearest sites of the
lattice. Other finite differences such as central one, $\left( \Delta
^{central}\right) f\left( x\right) =f\left( x+h/2\right) -f\left(
x-h/2\right) $ and the general eccentric one $\left( \Delta
^{eccentric}\right) f\left( x\right) =f\left( x+\left( h-\varepsilon \right)
\right) -f\left( x-\varepsilon \right) ,(\varepsilon \in \left( 0,h\right) $
and $\varepsilon \neq h/2)$ are also mathematically definable, but
irrelevant to our problems. For our functions are defined on the simple
cubic lattice whose points lie at positions ($x_{1},x_{2},x_{3},...,x_{k}$)
in the $k$-dimensional Cartesian space, where $x_{1},x_{2},x_{3},...,x_{k}$
are integers, and $h$ is the lattice constant which can be taken as unit, $%
h=1$, for convenience. In consequence, the second order \emph{differences}
belong to two groups respectively: the \emph{forward one}, $\left( \Delta
^{f}\right) ^{2}f\left( x\right) =f\left( x+2h\right) -f\left( x+h\right)
-\left( f\left( x+h\right) -f\left( x\right) \right) =f\left( x+2h\right)
-2f\left( x+h\right) +f\left( x\right) $, and the \emph{backward one} $%
\left( \Delta ^{b}\right) ^{2}f\left( x\right) =f\left( x\right) -f\left(
x-h\right) -\left( f\left( x-h\right) -f\left( x-2h\right) \right) =f\left(
x\right) -2f\left( x-h\right) +f\left( x-2h\right) $. A very important
simple function appears as $\left( \Delta ^{f}\right) f\left( x\right)
=\left( \Delta ^{b}\right) f\left( x\right) $, we denote these functions by 
\QTR{cal}{L}, which implies $l\left( x\right) =const.$, or $l\left( x\right) 
$ is linear in $x$ once $l\left( x\right) \in $\QTR{cal}{L}. In our
treatment, the meaningful function $l\left( x\right) $ can not be separated
from $f(x)$ which satisfies $\left( \Delta ^{b}\right) f\left( x\right) \neq
\left( \Delta ^{f}\right) f\left( x\right) $.

Secondly, we can introduce the sum of two functions $f\left( x\right)
=f_{1}\left( x\right) +f_{2}\left( x\right) $, and so forth, the sum of many
functions. Once $f\left( x\right) =f_{1}\left( x\right) +f_{2}\left(
x\right) $, a sum of two functions and either of them is purely an $l\left(
x\right) $, the first order of differences $\Delta f$ actually \emph{means}
four combinations: 
\begin{subequations}
\begin{eqnarray}
&&\left( \Delta ^{f}\right) f_{1}\left( x\right) +\left( \Delta ^{f}\right)
f_{2}\left( x\right) ,\text{denoted by 1f2f, group 1,}  \label{1-1} \\
&&\left( \Delta ^{f}\right) f_{1}\left( x\right) +\left( \Delta ^{b}\right)
f_{2}\left( x\right) ,\text{denoted by 1f2b, group 2,} \\
&&\left( \Delta ^{b}\right) f_{1}\left( x\right) +\left( \Delta ^{f}\right)
f_{2}\left( x\right) ,\text{denoted by 1b2f, group 3,} \\
&&\left( \Delta ^{b}\right) f_{1}\left( x\right) +\left( \Delta ^{b}\right)
f_{2}\left( x\right) ,\text{denoted by 1b2b, group 4.}  \label{1-4}
\end{eqnarray}%
These combinations can be grouped into four distinct types, the usual \emph{%
synchronous finite differences 1f2f and 1b2b, and }the\emph{\ asynchronous
finite differences 1f2b and 1b2f, which were simply overlooked before.}
Accordingly, the second order of differences $\left( \Delta \right)
^{2}f\left( x\right) $ for these four groups, 1f2f, 1f2b, 1b2f and 1b2b are
given by, respectively, 
\end{subequations}
\begin{subequations}
\begin{eqnarray}
&&\left( \Delta ^{f}\right) ^{2}f_{1}\left( x\right) +\left( \Delta
^{f}\right) ^{2}f_{2}\left( x\right) ,\text{for 1f2f,}  \label{2-1} \\
&&\left( \Delta ^{f}\right) ^{2}f_{1}\left( x\right) +\left( \Delta
^{b}\right) ^{2}f_{2}\left( x\right) ,\text{for 1f2b,} \\
&&\left( \Delta ^{b}\right) ^{2}f_{1}\left( x\right) +\left( \Delta
^{f}\right) ^{2}f_{2}\left( x\right) ,\text{for 1b2f,} \\
&&\left( \Delta ^{b}\right) ^{2}f_{1}\left( x\right) +\left( \Delta
^{b}\right) ^{2}f_{2}\left( x\right) ,\text{for 1b2b.}  \label{2-4}
\end{eqnarray}%
Similarly, we can define higher order of differences $\left( \Delta \right)
^{j}f\left( x\right) $ $\left( j=3,4,5,...\right) $ for these four groups,
but these cases do not interest us for the present.

Thirdly, we define the \emph{difference quotients of different orders. }For
a function, we\emph{\ }have \emph{forward difference quotients of different
orders }$\left( \Delta ^{f}\right) f\left( x\right) /h$, $\left( \Delta
^{f}\right) ^{2}f\left( x\right) /h^{2}$, ..., and \emph{backward difference
quotients of different orders}\ are, $\left( \Delta ^{b}\right) f\left(
x\right) /h$, $\left( \Delta ^{b}\right) ^{2}f\left( x\right) /h^{2}$, ...,
respectively. With $h=1$, we have $\left( \Delta ^{f}\right) ^{i}f\left(
x\right) /h^{i}=\left( \Delta ^{f}\right) ^{i}f\left( x\right) $ and $\left(
\Delta ^{b}\right) ^{i}f\left( x\right) /h^{i}=\left( \Delta ^{b}\right)
^{i}f\left( x\right) $ ($i=1,2,3,...$), and the \emph{difference quotients
of second order }for the sum of two functions $f\left( x\right) =f_{1}\left(
x\right) +f_{2}\left( x\right) $ are given\emph{\ }in Eqs. (\ref{2-1})-(\ref%
{2-4}).

Fourthly, we consider one discrete function $\Psi \left( n\right) $ defined,
for convenience, on the interval of semi-positive integers $n\in 
\mathbb{Z}
^{+}$, which has the local maxima and minima, subject to some equality
constraints $\mathbf{\psi =}\left( \psi _{1},\psi _{2},\psi _{3},...\right) =%
\mathbf{0}$ and the constraints belong to \QTR{cal}{L}. For finding the
local maxima and minima of the function $\Psi \left( n\right) $, we
construct a functional $\Phi $ which behaves as a function, 
\end{subequations}
\begin{equation}
\Phi =\Psi \left\{ n\right\} +\mathbf{\alpha \cdot \psi ,}  \label{Theorem}
\end{equation}%
where $\mathbf{\alpha =}\left( \alpha _{1},\alpha _{2},\alpha
_{3},...\right) $ are Lagrange multipliers each of which $\alpha _{i}$
companies a constraint condition $\psi _{i}=0$. The local maxima and minima
satisfy, 
\begin{equation}
\delta \Phi =0.  \label{variation}
\end{equation}%
Since the discreteness of the function $\Psi \left( n\right) $, the smallest
finite change of $n$ is $\Delta n=1$, and we have accordingly two
differences (or difference quotients) of the function $\Psi \left\{
n\right\} $ in the following:\ $\left( \delta ^{f}\right) \Psi \left\{
n\right\} /\delta n=\Psi \left\{ n+1\right\} -\Psi \left\{ n\right\} $ $%
=\left( \Delta ^{f}\right) \Psi \left\{ n\right\} $ for the forward
difference and $\left( \delta ^{b}\right) \Psi \left\{ n\right\} /\delta
n=\Psi \left\{ n\right\} -\Psi \left\{ n-1\right\} =\left( \Delta
^{b}\right) \Psi \left\{ n\right\} $ for the backward difference. In
consequence, two relations $n^{\mu }=n^{\mu }\left( \mathbf{\alpha }\right) $
($\mu =1,2$) solve the equation (\ref{variation}), which contains both
spurious and true one. Then how to determine the true one?

Assuming that all relations $n^{\mu }=n^{\mu }\left( \mathbf{\alpha }\right) 
$ have different values of the second order differences, and denoting $%
n^{1}=n^{1}\left( \mathbf{\alpha }\right) $ that solves the forward
difference variation as $\left( \Delta ^{f}\right) \Psi \left\{
n^{1}\right\} +\mathbf{\alpha \cdot }\Delta ^{f}\mathbf{\psi }\left\{
n^{1}\right\} =0$, the another solution $n^{2}=n^{2}\left( \mathbf{\alpha }%
\right) $ solves the backward difference variation as $\left( \Delta
^{b}\right) \Psi \left\{ n^{2}\right\} +\mathbf{\alpha \cdot }\Delta ^{b}%
\mathbf{\psi }\left\{ n^{2}\right\} =0$. If only only one solution leads to
the\ maximal distribution, the true solution is obtained, the simplest case.
Now we deal with complicated case. Assuming that $\left( \Delta ^{f}\right)
^{2}\Psi \left\{ n^{1}\right\} <0$ and $\left( \Delta ^{b}\right) ^{2}\Psi
\left\{ n^{2}\right\} \left( \neq \left( \Delta ^{f}\right) ^{2}\Psi \left\{
n^{1}\right\} \right) <0$, which amounts to assuming that both $n^{1}\left( 
\mathbf{\alpha }\right) \ $and $n^{2}\left( \mathbf{\alpha }\right) $ can
give maximal value for $\Psi \left( n\right) $. If $\left\vert \left( \Delta
^{f}\right) ^{2}\Psi \left\{ n^{1}\right\} \right\vert >\left\vert \left(
\Delta ^{b}\right) ^{2}\Psi \left\{ n^{2}\right\} \right\vert $, the
solution $n^{1}=n^{1}\left( \mathbf{\alpha }\right) $ is true, and \textit{%
vice versa}. In other words, the second order difference of the true
solution take largest value among two maximal values $\left\{ \left\vert
\left( \Delta \right) ^{2}\Psi \left\{ n^{1}\right\} \right\vert ,\left\vert
\left( \Delta \right) ^{2}\Psi \left\{ n^{2}\right\} \right\vert \right\} $.

Once the \emph{discrete function} $\Psi \left( n\right) $ is the sum of two
functions\emph{\ }$\Psi _{1}\left( n\right) $ and $\Psi _{2}\left( n\right) $%
, $\Psi \left( n\right) =\Psi _{1}\left( n\right) +\Psi _{2}\left( n\right) $
and either of them belongs to \QTR{cal}{L}, there are four relations $n^{\xi
}=n^{\xi }\left( \mathbf{\alpha }\right) $ ($\xi =1,2,3,4$), solving,
respectively, 
\begin{subequations}
\begin{eqnarray}
\left( \Delta ^{f}\right) \Psi _{1}\left( n^{1}\right) +\left( \Delta
^{f}\right) \Psi _{2}\left( n^{1}\right) +\mathbf{\alpha \cdot }\Delta 
\mathbf{\psi }\left\{ n^{1}\right\}  &=&0,\text{ for 1f2f,}  \label{T21} \\
\left( \Delta ^{f}\right) \Psi _{1}\left( n^{2}\right) +\left( \Delta
^{b}\right) \Psi _{2}\left( n^{2}\right) +\mathbf{\alpha \cdot }\Delta 
\mathbf{\psi }\left\{ n^{2}\right\}  &=&0,\text{ for 1f2b,}  \label{T22} \\
\left( \Delta ^{b}\right) \Psi _{1}\left( n^{3}\right) +\left( \Delta
^{f}\right) \Psi _{2}\left( n^{3}\right) +\mathbf{\alpha \cdot }\Delta 
\mathbf{\psi }\left\{ n^{3}\right\}  &=&0,\text{ for 1b2f,}  \label{T23} \\
\left( \Delta ^{b}\right) \Psi _{1}\left( n^{4}\right) +\left( \Delta
^{b}\right) \Psi _{2}\left( n^{4}\right) +\mathbf{\alpha \cdot }\Delta 
\mathbf{\psi }\left\{ n^{4}\right\}  &=&0,\text{ for 1b2b.}  \label{T24}
\end{eqnarray}%
In similar manner, we deal with only the complicated case. The true solution 
$n^{\xi }=n^{\xi }\left( \mathbf{\alpha }\right) $ must be that whose second
order difference $\Delta ^{2}\Psi \left( n^{\xi }\right) $ takes the largest
value in magnitude among all solutions from four groups, 1f2f, 1f2b, 1b2f
and 1b2b. I.e., 
\end{subequations}
\begin{equation}
\left\vert \Delta ^{2}\Psi \left( n^{\xi }\right) \right\vert =\max \left\{
\left\vert \Delta ^{2}\Psi \left\{ n^{1}\right\} \right\vert \text{,}%
\left\vert \Delta ^{2}\Psi \left\{ n^{2}\right\} \right\vert \text{,}%
\left\vert \Delta ^{2}\Psi \left\{ n^{3}\right\} \right\vert \text{,}%
\left\vert \Delta ^{2}\Psi \left\{ n^{4}\right\} \right\vert \right\} .
\label{Sol}
\end{equation}

Five immediate remarks follow. 1. In our procedure above, the \emph{correct
solution} and the \emph{true solution} differ; and the former refers to its
solving the first order equation (\ref{variation}) and the latter refers to
its maximizing the second order variations in magnitude. In contrast, the
conventional procedure accepts all solutions once they are stationary. 2.
Once the second order variations give infinite values in magnitude, the
corresponding solution is acceptable. 3. If there are solutions which have
no difference up to second differences such that we can not identify which
is true, higher order differences can be invoked; and our procedure can be
easily generalized for the \emph{discrete function} $\Psi \left( n\right) $
that is the sum of more functions. 4. If the constraint function $\mathbf{%
\psi }$ is nonlinear in $n$, the problem must be treated in similar manner.
5. Our procedure is quite general, not limited to method of MPD, but the
present application is limited to it.

It is worth to emphasize: a proper formalism of discrete calculus of
variations must be taken account of both the \emph{synchronous finite
differences and asynchronous finite differences, }allowing for solution
groups of finite differences based on \emph{all possible combinations of
forward and backward difference}. In the following section, I\ will
illustrate the fundamental importance of the \emph{asynchronous finite
differences} with a detailed derivation of the exact form of the Bose
distribution, and briefly discuss the Boltzmann and Fermi distribution.

\section{Asynchronous finite differences and exact distributions}

Considering a system of $N$ noninteracting, indistinguishable particles
confined to a space of volume $V$ and sharing a given energy $E$. Let $%
\varepsilon _{i}$ denote the energy of $i$-th level and $\varepsilon
_{1}\prec \varepsilon _{2}\prec \varepsilon _{3}\prec ...$, and $g_{i}$
denote the degeneracy of the level. In a particular situation, we may have $%
n_{1}$ particles in the first level $\varepsilon _{1}$, $n_{2}$ particles in
the second level $\varepsilon _{2}$, and so on, defining a distribution set $%
\left\{ n_{i}\right\} $. The number of the distinct microstates in set $%
\left\{ n_{i}\right\} $ is then given by,%
\begin{equation}
\Omega \left\{ n_{i}\right\} =\prod_{i}\frac{(n_{i}+g_{i}-1)!}{%
n_{i}!(g_{i}-1)!}.  \label{BOSE}
\end{equation}%
The Bose functional $f$ is with two Lagrange multipliers $\alpha $ and $%
\beta $,%
\begin{equation}
f=\sum_{i}\left( \ln (n_{i}+g_{i}-1)!-\ln n_{i}!-\ln (g_{i}-1)!\right)
-\alpha \left( \sum_{i}n_{i}-N\right) -\beta \left( \sum_{i}n_{i}\varepsilon
_{i}-E\right) .  \label{BOSE-2}
\end{equation}%
The variational $\delta f$ is, 
\begin{eqnarray}
\delta f &=&\sum_{i}\left\{ \delta \ln (n_{i}+g_{i}-1)!-\delta \ln
n_{i}!\right\} -\delta n_{i}\left( \alpha +\beta \varepsilon _{i}\right) 
\notag \\
&=&\sum_{i}\delta n_{i}\left\{ \frac{\delta \ln (n_{i}+g_{i}-1)!}{\delta
n_{i}}-\frac{\delta \ln n_{i}!}{\delta n_{i}}-\left( \alpha +\beta
\varepsilon _{i}\right) \right\} .  \label{BOSE-3}
\end{eqnarray}%
Since the independence of variables $n_{i}$, $\delta f=0$ leads to,%
\begin{equation}
\frac{\delta \ln (n_{i}+g_{i}-1)!}{\delta n_{i}}-\frac{\delta \ln n_{i}!}{%
\delta n_{i}}-\left( \alpha +\beta \varepsilon _{i}\right) =0.
\label{BOSE-4}
\end{equation}%
Thus, there are essentially two functions\emph{\ }$\Psi _{1}=\ln
(n_{i}+g_{i}-1)!$ and $\Psi _{2}=\ln n_{i}!$, and $-n_{i}\left( \alpha
+\beta \varepsilon _{i}\right) $ is clearly linear in $n_{i}$ and can be a
part of $\Psi _{1}$ or $\Psi _{1}$\emph{. }For $\ln (n_{i}+g_{i}-1)!$ and $%
\ln n_{i}!$. The smallest forward and backward differences are given in
Table \ref{T1}. Four combinations constructing $\Delta \ln
(n_{i}+g_{i}-1)!/\Delta n_{i}-\Delta \ln n_{i}!/\Delta n_{i}$ are summarized
in Table\ II. Accordingly, we have four solutions presented in Table III,
and all satisfy $\delta f=0$. These distributions are mutually different
when $n_{i}\sim 1$, though all of them converge to the same one when $%
n_{i}\gg 1$, 
\begin{equation}
n_{i}\approx \frac{g_{i}}{e^{\alpha +\beta \varepsilon _{i}}-1}.
\end{equation}%
\begin{table}[h]
\caption{The smallest forward/backward difference of $\Psi _{1}=\ln
(n_{i}+g_{i}-1)!$ and $\Psi _{2}=\ln n_{i}!$.}
\label{T1}\centering
\par
\begin{tabular}{|l|l|l|}
\hline
& $\Psi _{1}=\ln (n_{i}+g_{i}-1)!$ & $\Psi _{2}=\ln n_{i}!$ \\ \hline
forward & $\ln (n_{i}+g_{i})$ & $\ln (n_{i}+1)$ \\ \hline
backward & $\ln (n_{i}+g_{i}-1)$ & $\ln (n_{i})$ \\ \hline
\end{tabular}%
\end{table}
\begin{table}[h]
\caption{The column and row give the forward/backward differences for $\Psi
_{1}=\ln (n_{i}+g_{i}-1)!$ and $\Psi _{2}=\ln n_{i}!$, respectively, and we
list the results of four combinations $\Delta \ln (n_{i}+g_{i}-1)!-\Delta
\ln n_{i}!$.}
\label{T2}\centering
\par
\begin{tabular}{|l|l|l|}
\hline
& $\Psi _{2}$ forward & $\Psi _{2}$ backward \\ \hline
$\Psi _{1}$ forward & $\ln (n_{i}+g_{i})-\ln (n_{i}+1)$ for 1f2f & $\ln
(n_{i}+g_{i})-\ln n_{i}$ for 1f2b \\ \hline
$\Psi _{1}$ backward & $\ln (n_{i}+g_{i}-1)-\ln (n_{i}+1)$ for 1b2f & $\ln
(n_{i}+g_{i}-1)-\ln n_{i}$ for 1b2b \\ \hline
\end{tabular}%
\end{table}
\begin{table}[h]
\caption{Four solutions}
\label{T3}\centering
\par
\begin{tabular}{|l|l|l|}
\hline
& $\Psi _{2}$ forward & $\Psi _{2}$ backward \\ \hline
$\Psi _{1}$ forward & $\frac{n_{i}+g_{i}}{n_{i}+1}=\exp \left( \alpha +\beta
\varepsilon _{i}\right) $ for 1f2f & $\frac{n_{i}+g_{i}}{n_{i}}=$ $\exp
\left( \alpha +\beta \varepsilon _{i}\right) $ for 1f2b \\ \hline
$\Psi _{1}$ backward & $\frac{n_{i}+g_{i}-1}{n_{i}+1}=$ $\exp \left( \alpha
+\beta \varepsilon _{i}\right) $ for 1b2f & $\frac{n_{i}+g_{i}-1}{n_{i}}%
=\exp \left( \alpha +\beta \varepsilon _{i}\right) $ for 1b2b \\ \hline
\end{tabular}%
\end{table}
\begin{table}[h]
\caption{The column and row give the forward/backward finite differences for 
$\ln (n_{i}+g_{i}-1)!$ and $\ln n_{i}!$, respectively, and then forming $%
\Delta ^{2}\ln (n_{i}+g_{i}-1)!-\Delta ^{2}\ln n_{i}!$ accordingly.}
\label{T4}\centering
\par
\begin{tabular}{|l|l|l|}
\hline
& $\Psi _{2}$ forward & $\Psi _{2}$ backward \\ \hline
$\Psi _{1}$ forward & $-\ln (\frac{n_{i}+2}{n_{i}+1}\frac{n_{i}+g_{i}}{%
n_{i}+g_{i}+1})$\ for 1f2f & $-\ln (\frac{n_{i}}{n_{i}-1}\frac{n_{i}+g_{i}}{%
n_{i}+g_{i}+1})$\ for 1f2b \\ \hline
$\Psi _{1}$ backward & $-\ln (\frac{n_{i}+2}{n_{i}+1}\frac{n_{i}+g_{i}-2}{%
n_{i}+g_{i}-1})$ for 1b2f & $-\ln (\frac{n_{i}}{n_{i}-1}\frac{n_{i}+g_{i}-2}{%
n_{i}+g_{i}-1})$\ for 1b2b \\ \hline
\end{tabular}%
\end{table}
The second order variations of $\delta ^{2}\ln \Omega \left\{
n_{i}{}\right\} $ for the four solutions are explicitly shown in the Table
IV, which are crucial for us to identify the true solution among all
possible ones in Table III. It is easily to verify that one combination 1f2b
is right for we have, from results in Table IV, 
\begin{subequations}
\begin{eqnarray}
\frac{n_{i}}{n_{i}-1}\frac{n_{i}+g_{i}}{n_{i}+g_{i}+1}-\frac{n_{i}}{n_{i}-1}%
\frac{n_{i}+g_{i}-2}{n_{i}+g_{i}-1} &=&\frac{n_{i}}{n_{i}-1}\left( \frac{2}{%
\left( n_{i}+g_{i}\right) ^{2}-1}\right) >0,\text{ }\left( n_{i}>1\right)
\label{Bose1} \\
\frac{n_{i}}{n_{i}-1}\frac{n_{i}+g_{i}}{n_{i}+g_{i}+1}-\frac{n_{i}+2}{n_{i}+1%
}\frac{n_{i}+g_{i}}{n_{i}+g_{i}+1} &=&\frac{n_{i}+g_{i}}{n_{i}+g_{i}+1}%
\left( \frac{2}{n_{i}{}^{2}-1}\right) >0,\text{ }\left( n_{i}>1\right)
\label{Bose2} \\
\frac{n_{i}}{n_{i}-1}\frac{n_{i}+g_{i}}{n_{i}+g_{i}+1}-\frac{n_{i}+2}{n_{i}+1%
}\frac{n_{i}+g_{i}-2}{n_{i}+g_{i}-1} &=&2\frac{%
(2n_{i}{}^{2}+2g_{i}n_{i}-2-g_{i}+g_{i}^{2})}{\left( \left(
n_{i}+g_{i}\right) ^{2}-1\right) \left( n_{i}{}^{2}-1\right) }>0,\text{ }%
\left( n_{i}>1\right) .  \label{Bose3}
\end{eqnarray}%
To note that these criteria hold only when $n_{i}\geq 2$, and the
distributions with $n_{i}=1$ must be studied with special care, which will
be done in next section. So far, we reproduce the Bose distribution with $%
n_{i}\geq 2$, nevertheless, 
\end{subequations}
\begin{equation}
n_{i}=\frac{g_{i}}{e^{\alpha +\beta \varepsilon _{i}}-1},\text{ }\left(
n_{i}\geq 2\right) .  \label{BED}
\end{equation}

Next, let us briefly consider Boltzmann and Fermi system. For the Boltzmann
system, the number of the distinct microstates in set $\left\{ n_{i}\right\} 
$ is then given by,%
\begin{equation}
\Omega \left\{ n_{i}\right\} =\prod_{i}\frac{\left( g_{i}\right) ^{n_{i}}}{%
n_{i}!}.
\end{equation}%
The variational calculation of $\delta f$ is with two Lagrange multipliers $%
\alpha $ and $\beta $, 
\begin{equation}
\delta f=\sum_{i}\left\{ \delta n_{i}\left( \ln g_{i}-\alpha -\beta
\varepsilon _{i}\right) -\delta \ln n_{i}!\right\} =\sum_{i}\left( \delta
n_{i}\left( \ln g_{i}-\alpha -\beta \varepsilon _{i}\right) -\delta \ln
n_{i}!\right) .  \label{vb}
\end{equation}%
Due to the independence of variables $n_{i}$, the first order variational
equation $\delta f=0$ amounts to following single equation, 
\begin{equation}
\frac{\delta f}{\delta n_{i}}=\left( \ln g_{i}-\alpha -\beta \varepsilon
_{i}\right) -\frac{\delta \ln n_{i}!}{\delta n_{i}}=0,\text{ i.e., }\frac{%
\delta \ln n_{i}!}{\delta n_{i}}=\ln g_{i}-\alpha -\beta \varepsilon _{i},
\label{Boltzmann}
\end{equation}%
in which there is one function $\ln n!$\emph{.}

Through the second order variationals, we can easily find that the exact
distribution for the Boltzmann system $n_{i}\geq 1$ appears at backward
difference of the function $\ln n!$ as $\left( \Delta ^{b}\right) \ln n!=\ln
n_{i}$ in $\delta \ln n_{i}!/\delta n_{i}$ in Eq. (\ref{Boltzmann}), and the
result is, 
\begin{equation}
n_{i}=g_{i}e^{-\left( \alpha +\beta \varepsilon _{i}\right) },\text{ }\left(
n_{i}\geq 1\right) ,  \label{BD}
\end{equation}%
which takes the same form as the conventional one.

For the Fermi system, the number of the distinct microstates in set $\left\{
n_{i}\right\} $ is then given by,%
\begin{equation}
\Omega \left\{ n_{i}\right\} =\prod_{i}\frac{g_{i}!}{n_{i}!(g_{i}-n_{i})!}.
\end{equation}%
The variational calculation of $\delta f$ is with two Lagrange multipliers $%
\alpha $ and $\beta $, 
\begin{equation}
\delta f=-\sum_{i}\left\{ \delta \ln n_{i}!+\delta \ln
(g_{i}-n_{i})!\right\} +\delta n_{i}\left( \alpha +\beta \varepsilon
_{i}\right) .
\end{equation}%
Due to the independence of variables $n_{i}$, the first order variational
equation $\delta f=0$ amounts to following single equation, 
\begin{equation}
\frac{\delta \ln (g_{i}-n_{i})!}{\delta n_{i}}+\frac{\delta \ln n_{i}!}{%
\delta n_{i}}+\left( \alpha +\beta \varepsilon _{i}\right) =0,\text{i.e., }%
\frac{\delta \ln (g_{i}-n_{i})!}{\delta n_{i}}+\frac{\delta \ln n_{i}!}{%
\delta n_{i}}=-\alpha -\beta \varepsilon _{i},  \label{fermi}
\end{equation}%
in which there are two functions, and the first is $\ln (g_{i}-n_{i})!$ and
the second is $\ln n!$.

Also through the second order variationals, the exact distribution for the
Fermi system $n_{i}\geq 2$ appears at $\left( \Delta ^{f}\right) \ln
(g_{i}-n_{i})!+\left( \Delta ^{b}\right) \ln n!=-\ln (g_{i}-n_{i})+\ln n_{i}$
in $\delta \ln (g_{i}-n_{i})!/\delta n_{i}+\delta \ln n_{i}!/\delta n_{i}$ (%
\ref{fermi}), denoted by 1f2b, and the result is, 
\begin{equation}
n_{i}=\frac{g_{i}}{e^{\alpha +\beta \varepsilon _{i}}+1},\text{ }\left(
n_{i}\geq 2\right) ,  \label{FD}
\end{equation}%
which takes the same form as the conventional one, as well.

Clearly, our procedure puts no requirement on the total number of particle $%
N $ to be very large, though $n_{i}\geq 2$ is for Bose and Fermi system.
However, the distributions with $n_{i}=1$ for Fermi and Bose system can
actually be obtained in similar manner, which will be discussed in next
section.

\section{Exact distributions with finite numbers of particles}

The distributions with $n_{i}=0$ are physically irrelevant. This is because
two Lagrange multipliers $\alpha $ and $\beta $ are determined by two
constraints,%
\begin{equation}
N=\sum_{i=0}^{i_{\max }}n_{i},\text{ }E=\sum_{i=0}^{i_{\max
}}n_{i}\varepsilon _{i},  \label{LM}
\end{equation}%
in which there is no position for a term with $n_{i}=0$. In general, we have 
$n_{0}\geq n_{1}\geq ...\geq n_{i_{\max }}$ from distributions (\ref{BED}), (%
\ref{BD}), and (\ref{FD}). It is not necessary to assume $n_{i_{\max }}=1$,
but it is convenient to assume so then its explicit dependence on ($\alpha
,\beta ,\varepsilon _{i},g_{i}$) is what we want to obtain. Introducing the
probability $\rho _{i}$ via,%
\begin{equation}
\rho _{i}\equiv \frac{n_{i}}{N}=\frac{n_{i}}{\sum_{i=0}^{i_{\max }}n_{i}},
\end{equation}%
we can prove that $n_{i_{\max }}=1$ is physically insignificant because its
probability is completely ignorable due to the extremely high energy level.

First, let us deal with the possible exact distribution for Bose system with 
$n_{i_{\max }}=1$, from Eqs. (\ref{BOSE})-(\ref{BOSE-4}). The first and
second order variations of four solutions groups, and the results are
presented in Table V. 
\begin{table}[h]
\caption{Results for second order variations $\Delta ^{2}\ln
(n_{i}+g_{i}-1)!-\Delta ^{2}\ln n_{i}!$ for $n_{i}=0$ and $n_{i}=1$,
respectively. The results with $n_{i}=0$ are practically useless, and are
given only for reference. For $n_{i}=0$ second order variation runs into $%
\ln (-1)$\ that is undefinable within the real numbers dormain for both
solutions, 1f2b and 1b2b, and we can not judge whether these two solutions
are true or not.}\centering
\par
\begin{tabular}{|l|l|l|}
\hline
& $\delta f=0$ & $\delta ^{2}f$ \\ \hline
1f2f & $1=\frac{g_{i_{\max }}}{2\exp \left( \alpha +\beta \varepsilon
_{i_{\max }}\right) -1}$\  & $-\ln (\frac{3}{2}\frac{g_{i_{\max }}+1}{g_{i}+2%
})$\  \\ \hline
1f2b & $1=\frac{g_{i_{\max }}}{\exp \left( \alpha +\beta \varepsilon
_{i_{\max }}\right) -1}$ & $-\infty $ \\ \hline
1b2f & $1=\frac{g_{i_{\max }}}{2\exp \left( \alpha +\beta \varepsilon
_{i_{\max }}\right) }$ & $-\ln (\frac{3}{2}\frac{g_{i_{\max }}-1}{g_{i_{\max
}}})$ \\ \hline
1b2b & $1=\frac{g_{i_{\max }}}{\exp \left( \alpha +\beta \varepsilon
_{i_{\max }}\right) }$ & $-\infty $ \\ \hline
\end{tabular}%
\end{table}
Clearly, there are two solutions, 1f2b and 1b2b, that are satisfactory for
the second order variations are negatively infinitely large. Two solutions
from the 1f2b and 1b2b with $n_{i}=1$ are, respectively, from the results in
Table V, 
\begin{equation}
1=\frac{g_{i_{\max }}}{\exp \left( \alpha +\beta \varepsilon _{i_{\max
}}\right) -\delta }\text{, }\delta =1\text{ for 1f2b, and }\delta =0\text{
for 1b2b.}
\end{equation}%
It appears to be an arbitrariness in my procedure for there are two possible
solutions, 1f2b and 1b2b, and it seems to be no principle to exclude one of
them. However, there must be two solutions to account for two different
situations. On one hand, if there is only one particle in the system given $%
\alpha $ and $\beta $, no body can tell it is a Fermion or a Boson for there
is no other particle to exchange with it within the system. In consequence,
the distribution function must be Boltzmannian (\ref{BD}), which is what the
solution 1b2b implies. On the other hand, if there are other particles in
other energy levels, which can exchange with the particle in the highmost
energy level $i_{\max }$ and $n_{i_{\max }}=1$, the particle can be
identified as a Boson. In consequence, we must take an non-Boltzmannian
distribution, i.e., we are forced to choose another solution 1f2b in Table
V, 
\begin{equation}
n_{i}=\frac{g_{i}}{e^{\alpha +\beta \varepsilon _{i}}-1},\text{ }\left(
n_{i}\geq 1\right) .
\end{equation}%
It is the exact form of the Bose distribution.

In similar manner, we can deal with Fermi system with $n_{i}=1$, and we can
also take usual distribution, 
\begin{equation}
n_{i}=\frac{g_{i}}{e^{\alpha +\beta \varepsilon _{i}}+1},\text{ }\left(
n_{i}\geq 1\right) ,
\end{equation}%
as the exact form of the Fermi distribution.

Together with the exact form of the Boltzmann distribution (\ref{BD}), we
can conclude that the usual forms of the distribution functions for
Boltzmann, Bose and Fermi system are actually the exact ones. However, there
is a fundamental difference between our procedure and the usual one is that
ours is applicable to finite number of particles while the usual one holds
for very large number of particles. Clearly, when there is only one particle
in the system which keeps thermal contact with the heat bath and also keeps
averagely one particle in it, the particle obeys the Boltzmann distribution. 

\section{Conclusions and discussions}

In contrast to the continuous calculus of variations, the discrete one
possesses some peculiarities. A new discrete calculus is developed such that
the finite difference of the discrete function can not be treated in the 
\emph{synchronously} forward or backward, but must exhaust all possible
combinations. By use of the new discrete calculus, we carefully examine the
statistical distributions for Bose, Fermi and Boltzmann system, and
demonstrate that usual form of the distribution functions holds true exactly
for the number of particles is large than one. When there is only one
particle in it, there is no way to distinguish it as Boson or Fermion, our
theory automatically presents a result that the distribution is the
Boltzmann one, which is also compatible with the recent experiment on the
single particle engine. 

\begin{acknowledgments}
This work is financially supported by National Natural Science Foundation of
China under Grant No. 11675051.
\end{acknowledgments}

\end{document}